\documentclass[final,1p,times]{elsarticle} % do not change
\usepackage{graphicx} % do not change
\usepackage{amssymb} % do not change
\usepackage{amsthm} % do not change
\usepackage{lineno} % do not change

\journal{Nuclear Physics A} % do not change
\begin{document} % do not change

\begin{frontmatter} % do not change

%% QM09Author: please enter your  
%% Title, author and address info here; please do not use footnotes

% Your Title - please modify
\title{$\rho^0$ Production in Cu+Cu Collisions at $\sqrt{s_{NN}} =
200$ and 62.4 GeV in STAR}

% Principle author, and co-authors - please modify
\author{Patricia Fachini$^{a}$ for the STAR collaboration}

% Address - please modify
% note that if you have authors from several institutions, we recommend
% labelling these [a], [b], [c] etc.
\address[a]{Brookhaven National Laboratory, % label [a]
Bldg. 510 A,
Upton, NY, 11973, USA}

\begin{abstract} % do not change
%% Text of abstract goes here - please modify
The results on $\rho(770)^0$ production
in  Cu+Cu collisions at $\sqrt{s_{NN}} =
200$ and 62.4 GeV in STAR are
presented. The $\rho^0$ is measured via its hadronic decay
channel and used as a sensitive tool to examine the collision
dynamics in the hadronic medium. \end{abstract} % do not change

\end{frontmatter} % do not change

%% QM09: we keep linenumbers at least for initial version
%\linenumbers % do not change

%% start of main text - please modify. Below is a sub-set (single section) 
%% of an earlier proceedings that show how one can handle references 
%% and figures etc.
%%\section{}\label{}

\section{Introduction}
The study of the $\rho^0$ vector meson in relativistic heavy-ion collisions provides
information on the properties of the hot and dense medium created in such collisions. 
The $\rho^0$ measured via its hadronic decay channel can be used as a sensitive
tool to examine the collision dynamics in the hadronic medium through its decay and regeneration. 

The resonances that decay before kinetic freeze-out may not be
reconstructed due to the rescattering of the daughter particles.
In this case, the resonance survival probability is relevant and
depends on the time between chemical and kinetic freeze-outs, the
source size, and the $p_T$ of the resonance. On the other hand,
after chemical freeze-out, elastic interactions may increase the
resonance population compensating for the ones that decay before
kinetic freeze-out. This resonance regeneration depends on the
hadronic cross-section of their daughters. For example, the $K^*$
regeneration depends on the $K\pi$ cross section ($\sigma$) while
the rescattering of the daughter particles depends on
$\sigma_{\pi\pi}$ and $\sigma_{\pi p}$, which are 
larger (factor $\sim$3) than the $\sigma_{K\pi}$ \cite{13,14,15}. In
the case of the $\rho^0$, the regeneration probability that depends on the $\sigma_{\pi\pi}$ is
expected to be of the same order of the rescattering of the daughters. Thus, the study
of resonances can provide an independent probe of the time
evolution of the source from chemical to kinetic freeze-outs and
yield detailed information on hadronic interaction at later
stages.

The measurement of the elliptic flow ($v_2$) of hadrons shows that the $v_2$ scales 
with the number of constituent quarks. It has been proposed that the measurement of the 
$v_2$ of resonances can distinguish whether the resonance was produced at hadronization 
via quark coalescence or later in the collision via hadron re-scattering \cite{non}. The measurement of
the $\rho^0$ $v_2$ can potentially provide information on the $\rho^0$ production mechanism. 

\section{Results}
The $\rho(770)^0$ production corresponding to $20-60\%$ 
of the hadronic cross-section was measured via its
hadronic decay channel at midrapidity ($|y| \!\leq\! 0.5$) in
Cu+Cu collisions at $\sqrt{s_{_{NN}}}=$ 200 and 62.4 GeV
using the STAR detector at RHIC. In this analysis, the small signal to 
background ratio ($\sim 1/1000$) prevents the 
measurement of the $\rho^0$ in central Cu+Cu collisions.

\begin{figure}[ht]
\centering
\includegraphics[scale=0.41]{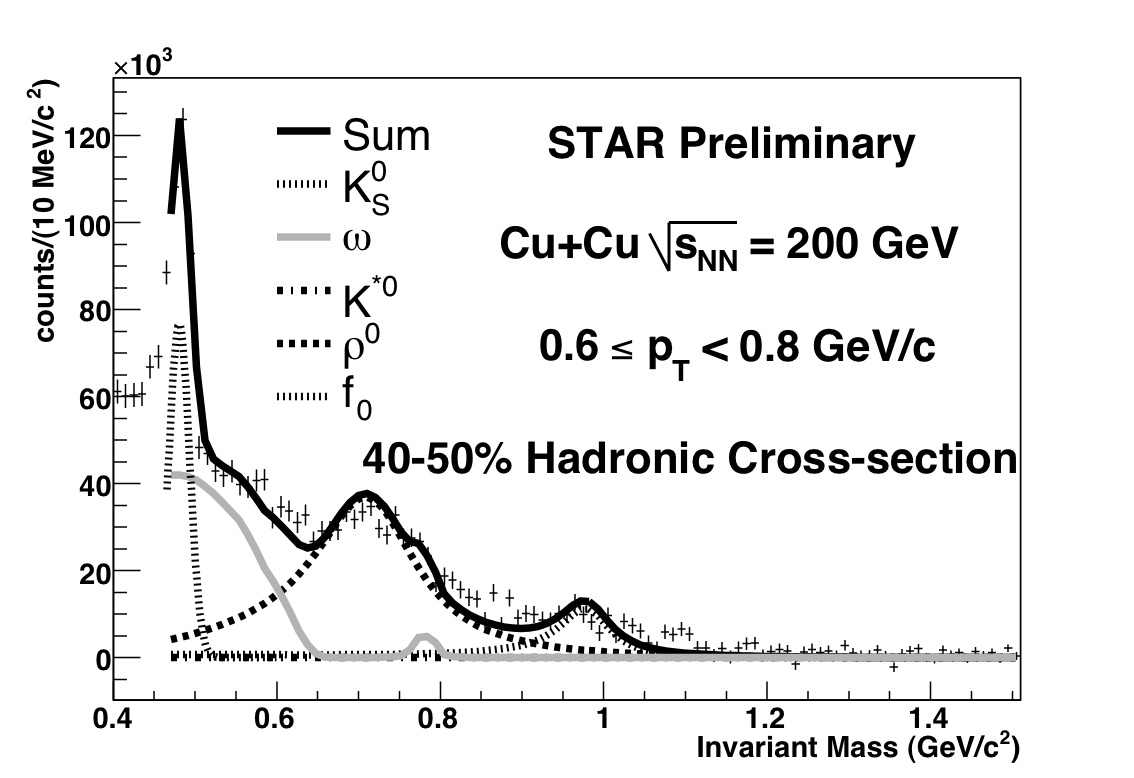}
\caption[]{The $\pi^+\pi^-$ invariant mass distributions after background
subtraction for a particular centrality of the hadronic Cu+Cu cross-section.}
\label{cocktail}
\end{figure}

The $\pi^+\pi^-$ invariant mass distributions after background
subtraction for a particular centrality of the hadronic Cu+Cu cross-section 
is shown in Fig. \ref{cocktail}.  The solid black line in Fig.~\ref{cocktail} is the sum of all
the contributions in the hadronic cocktail. The $K_S^0$ was fit to
a Gaussian (dotted line). The $\omega$ (light grey line) and
$K^{\ast}(892)^{0}$ (dash-dotted line) shapes were obtained from
the HIJING event generator \cite{26}, with the kaon being
misidentified as a pion in the case of the $K^*$. The
$\rho^0(770)$ (dashed line) and the $f_0(980)$ (dotted line) were fit by relativistic Breit-Wigner
functions times the phase space function described in \cite{16}. The masses of $K_S^0$,
$\rho^0$ and $f_0$ were free parameters in the fit, and
the widths of the $\rho^0$ and  $f_0$ were fixed according to the values in the PDG
\cite{pdg}. The uncorrected yields of $K_S^0$, $\rho^0$, $\omega$,
$f_0$, and $f_2$ were free parameters in the fit while the
$K^{\ast 0}$ fraction was fixed according to the
$K^{\ast}(892)^{0} \!\rightarrow\! \pi K$ measurement. The
$\rho^0$, $\omega$, $K^{\ast 0}$, $f_0$, and $f_2$ distributions
were corrected for the detector acceptance and efficiency
determined from a detailed simulation of the detector response.
This measurement does not have sufficient sensitivity to permit a
systematic study of the $\rho^0$ width. 
%Therefore, for the
%cocktail fits in this analysis, the $\rho^0$ width was fixed according to \cite{pdg}.

\begin{figure}[ht]
\centering
\includegraphics[width=0.46\textwidth]{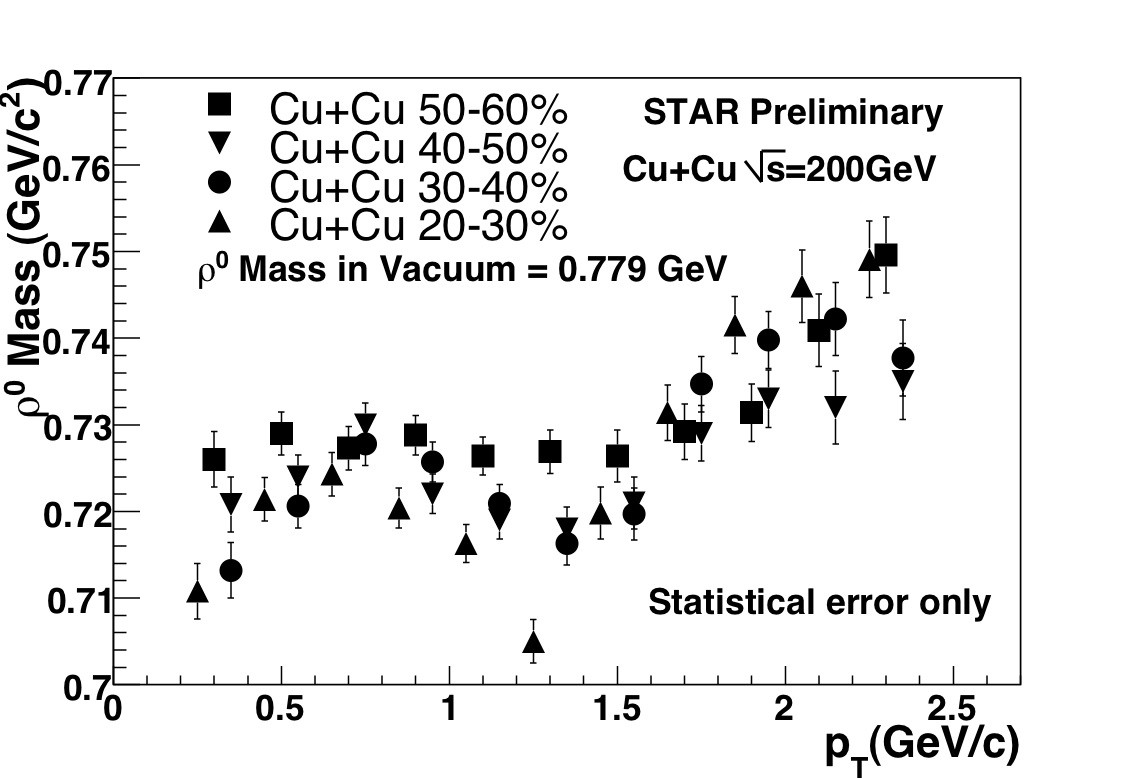}
\includegraphics[width=0.46\textwidth]{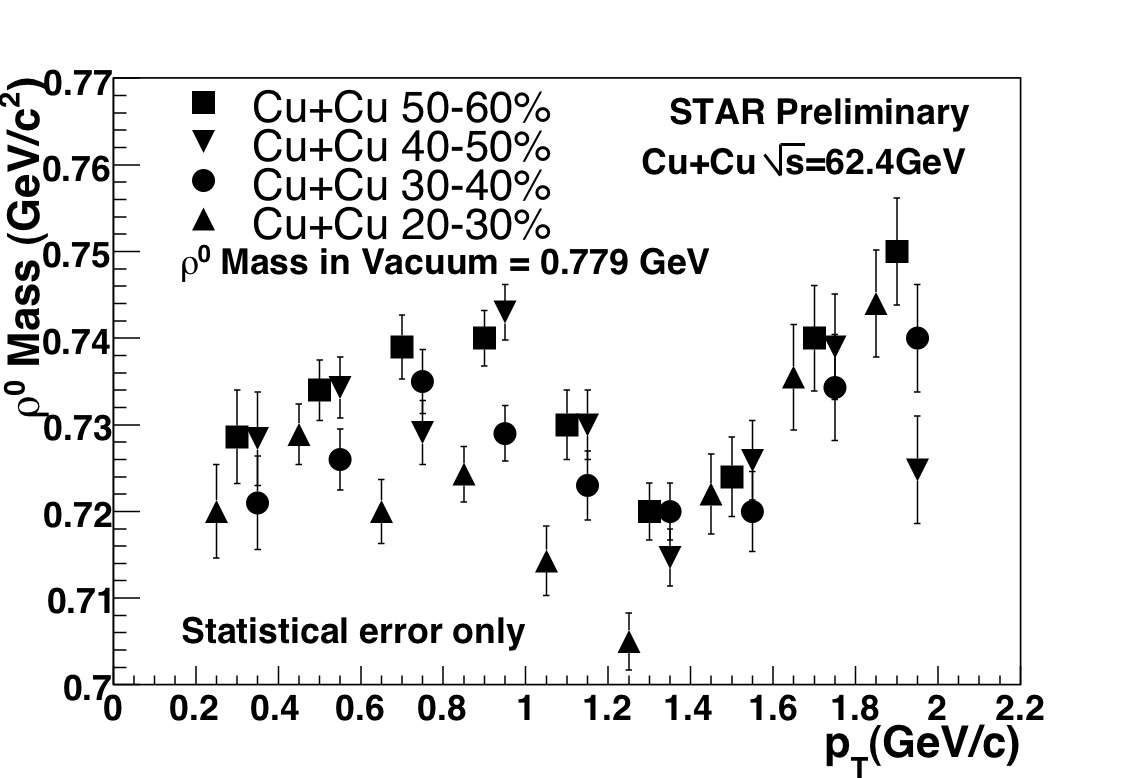}
\caption[]{The $\rho^0$ mass as a function of $p_T$ for 
$\sqrt{s_{_{NN}}} = $ 200 (left) and 62.4 (right) GeV.}
\label{mass}
\end{figure}

The $\rho^0$ mass is shown as a function of $p_T$ in
Fig.~\ref{mass} for four different centralities. The $\rho^0$ mass  seems to increase as a function of $p_T$
and is systematically lower than the value reported by the PDG of 775.49 $\pm$ 0.34 MeV/$c^2$\cite{pdg}. A mass shift 
has also been observed in $e^+e^-$, $p+p$ and peripheral Au+Au collisions, as described in \cite{16}.
Dynamical interactions with the surrounding matter, interference between various $\pi^+\pi^-$ scattering channels, phase 
space distortions due to the rescattering of pions forming $\rho^0$, and Bose-Einstein correlations between $\rho^0$ decay 
daughters and pions in the surrounding matter are possible 
explanations for the apparent modiÞcation of the $\rho^0$ 
meson properties \cite{16}.

\begin{figure}[ht]
\centering
\includegraphics[width=0.46\textwidth]{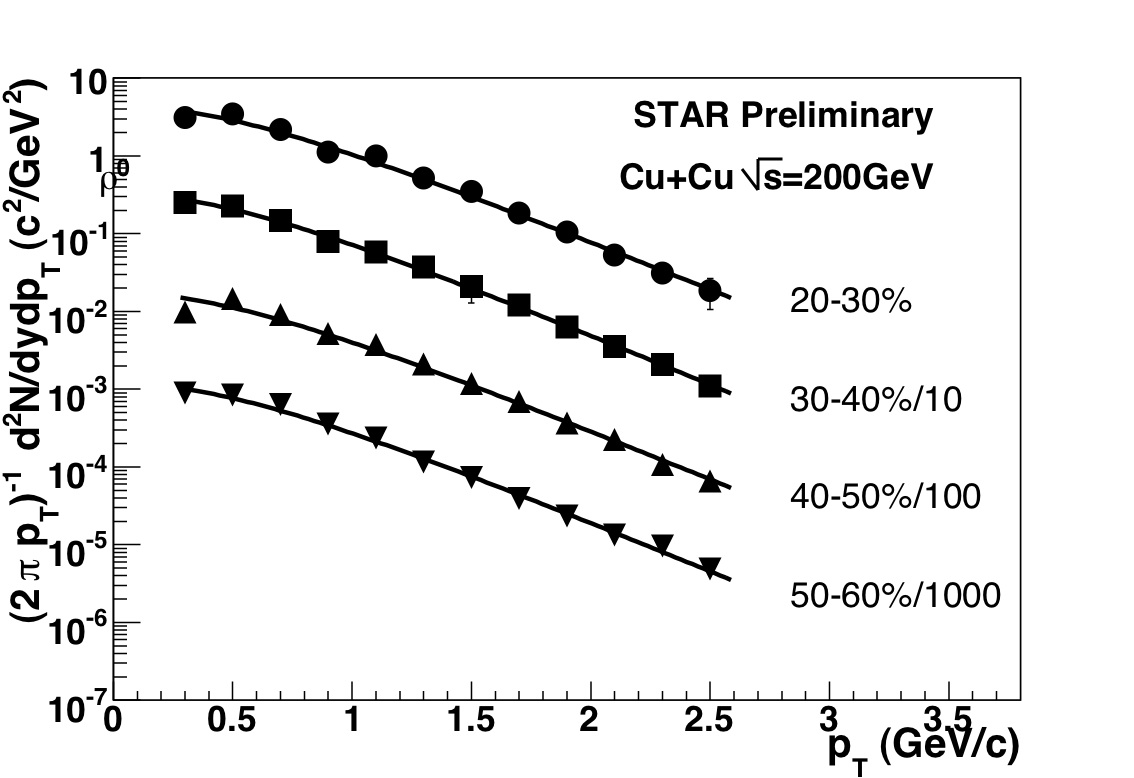}
\includegraphics[width=0.46\textwidth]{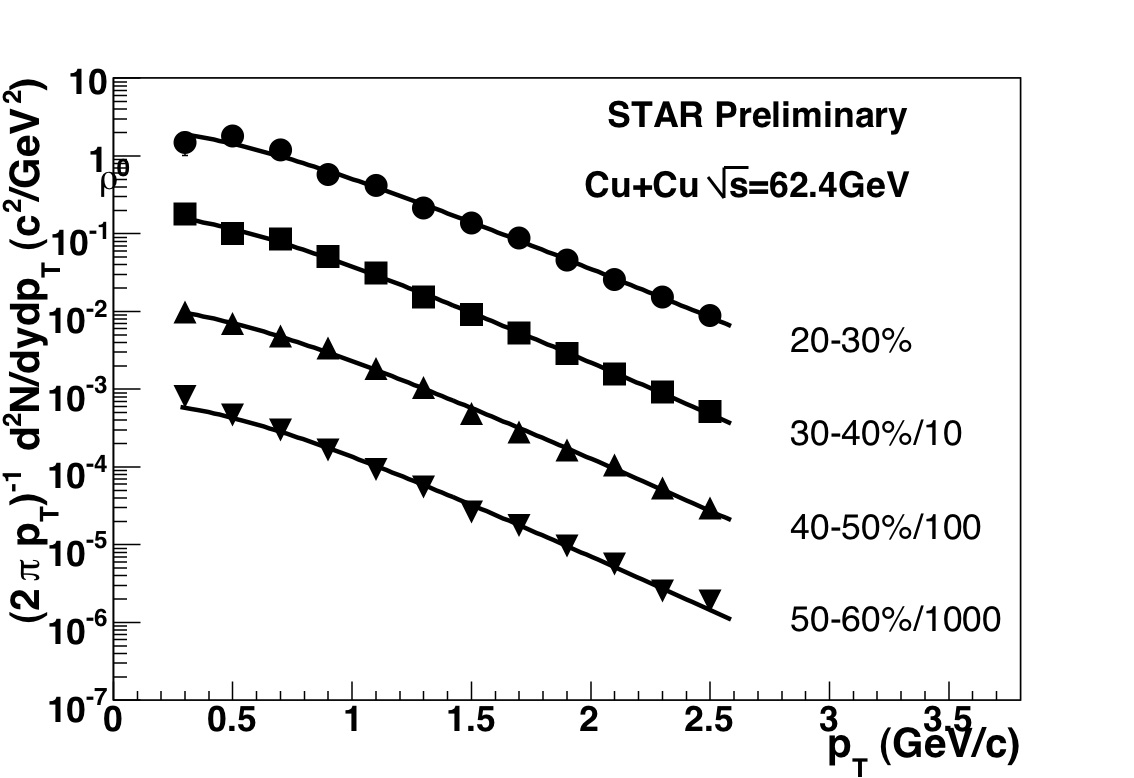}
\caption[]{The $\rho^0$ spectra as a function of $p_T$ for 
$\sqrt{s_{_{NN}}} = $ 200 (left) and 62.4 (right) GeV.}
\label{spectra}
\end{figure}

The corrected invariant yields [$d^2N/(2\pi p_Tdp_Tdy)$] at $|y|
\!<\!$ 0.5 as a function of $p_T$ for Cu+Cu collisions at 
$\sqrt{s_{_{NN}}} = $ 200 and 
62.4 GeV for four different centralities are shown in Fig.~\ref{spectra}. 
An exponential fit was used to extract the
$\rho^0$ yield per unit of rapidity around midrapidity. 

\begin{figure}[ht]
\centering
\includegraphics[width=0.46\textwidth]{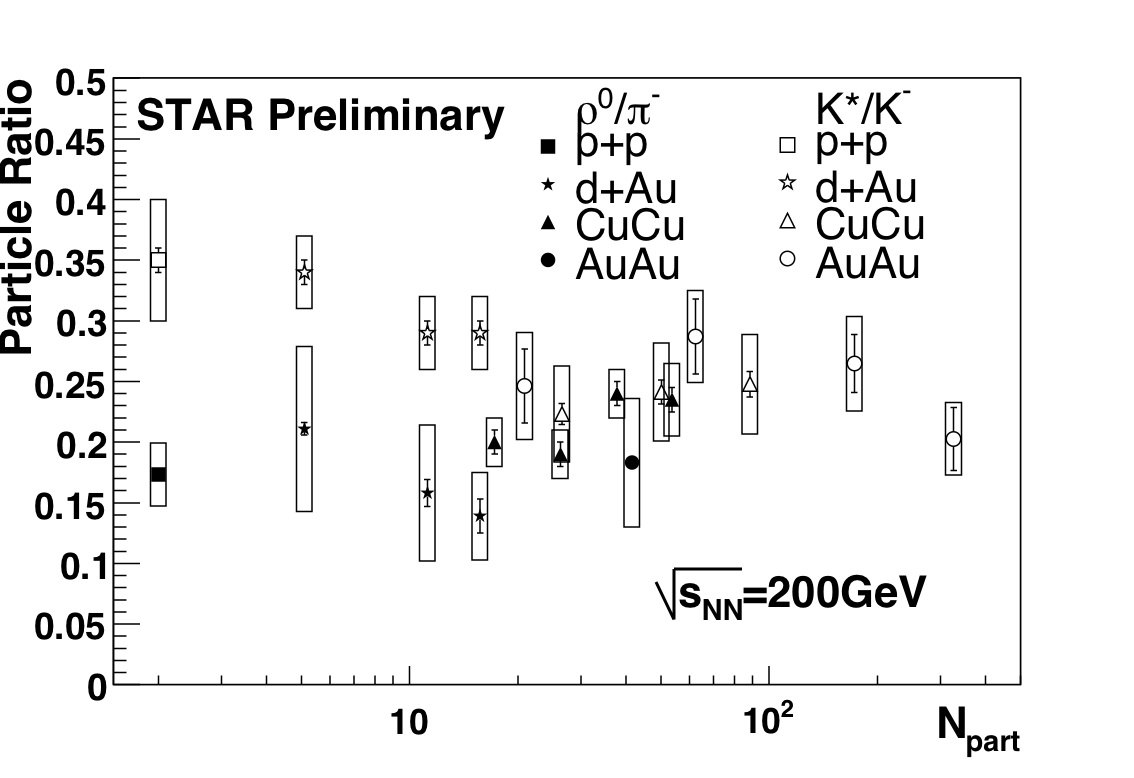}
\includegraphics[width=0.46\textwidth]{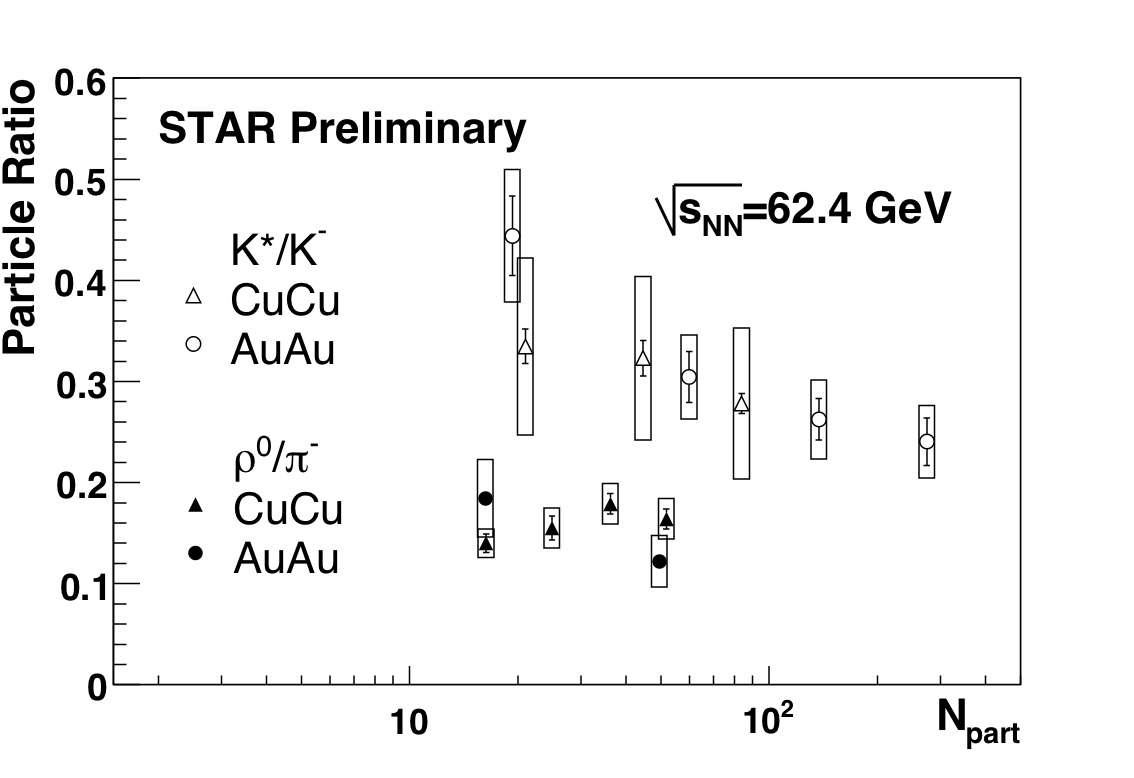}
\caption[]{Particle ratios as a function of the number of participants for 
$\sqrt{s_{_{NN}}}=$ 200 (left) and 62.4 (right) GeV.}
\label{ratios}
\end{figure}

The $\rho^0/\pi^-$ ratio together with the $K^*/K^-$ ratio are plotted 
in Fig.~\ref{ratios} as a function 
of the number of participants ($N_{part}$) 
for $\sqrt{s_{_{NN}}}=$ 200 and 62.4 GeV \cite{prc,dau}.  The $\rho^0/\pi^-$ and the $K^*/K^-$
ratios behave differently as a function of $N_{part}$.  The $K^*/K^-$ ratio decreases 
with $N_{part}$, while the $\rho^0/\pi^-$ ratio is independent or slightly 
increases with $N_{part}$. This can be understood in terms of hadron cross-sections ($\sigma$). 
The $\sigma_{K\pi} $ is smaller than $\sigma_{\pi\pi}$. Therefore, it is more probable for a pion 
to scatter with another pion and form a $\rho^0$ than it is for a kaon to scatter with a pion and form a $K^*$.
So, in the case of the $K^*$, the rescattering of the daughters will destroy the $K^*$. However,
in the case of the $\rho^0$, the rescattering of the daughters will most probably generate another $\rho^0$ 
and the regeneration will compensate the rescattering of the daughters and even increasing the number of 
$\rho^0$.

\begin{figure}[ht]
\centering
\includegraphics[scale=0.4]{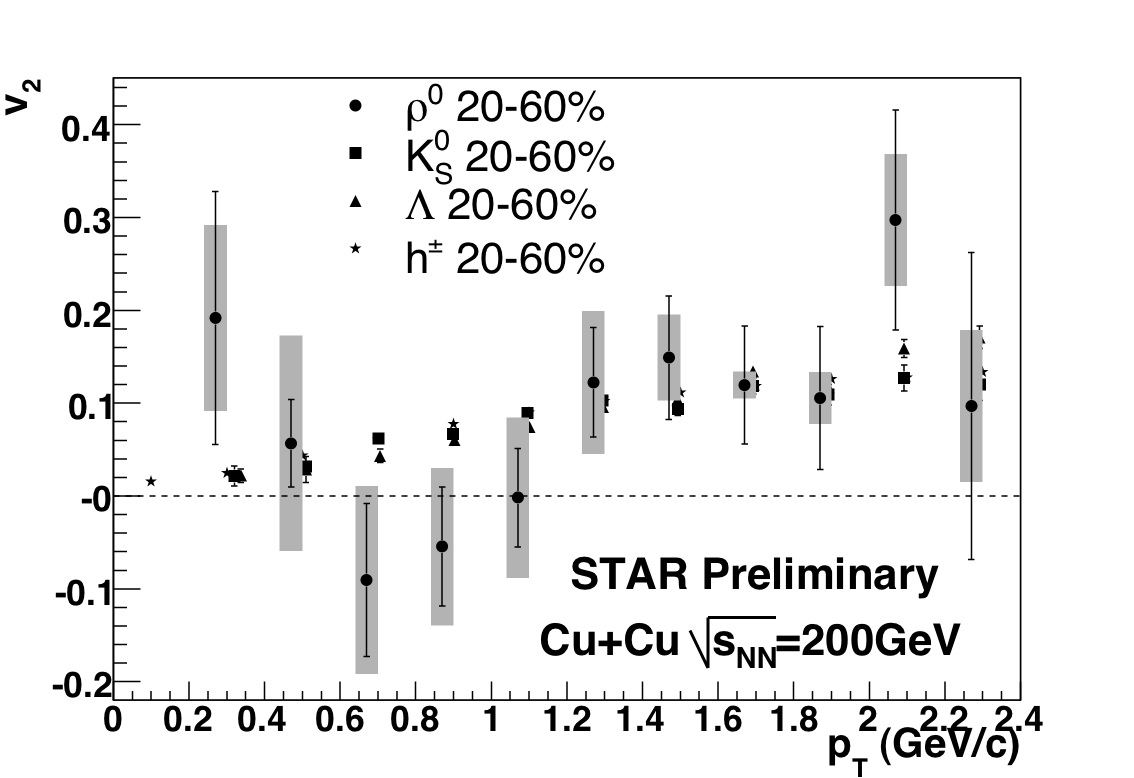}
\caption[]{$\rho^0$ $v_2$ measured in Cu+Cu collisions at $\sqrt{s_{_{NN}}} = $ 200 GeV.}
\label{v2}
\end{figure}

The $\rho^0$ $v_2$ as a function of $p_T$ is  shown in Fig.~\ref{v2}. In the same figure 
it is also plotted the $K^0_S$, $\Lambda$ and charged hadron $v_2$ for comparison, where 
all the measurements are from the same hadronic cross section \cite{v2}.  
We observe a non-zero $\rho^0$ $v_2$ of about 13 $\pm 3\%$ for $p_T > $ 1.2 GeV/$c$.
In order to calculate the contributions to the $\rho^0$ production from either direct quark
or hadron combinations, we use the function proposed in \cite{fit} to fit the $\rho^0$ $v_2$ distribution 
(0.3 $\leq p_T \leq$ 2.3 GeV/$c$)
and we obtain $n = 4.7 \pm 2.9$, where $n$ is a free parameter standing for the number of constituent 
quarks and the $\chi^2$/ $ndf$ = 10.3/9. Only the statistical error was considered in the fit.
 Due to the large statistics and the limited $p_T$ range measured, it is difficult to 
identify the $\rho^0$ production fractions from direct quark ($n = 2$) or hadron combinations ($n = 4$).  

%% end of main text

\section{Conclusions}

We measure significant $\rho^0$ production in Cu+Cu collisions 
at at $\sqrt{s_{_{NN}}} = $ 200 and 62 GeV and the production measured 
corresponds to $20-60\%$ of the hadronic cross-section. We observe a mass shift of about  
-45 MeV/$c^2$ possibly due to medium modifications, phase space, interference and  Bose-Einstein correlations. 
The rescattering of the daughters and the regeneration of the resonances are 
driven by the hadron cross-sections where the $K^*$ is dominated by the rescattering of its daughters 
and in the case of the $\rho^0$ the regeneration is compensating for the rescattering of the daughters. 
We presented the first measurement of the $\rho^0$ $v_2$ showing that the $\rho^0$  
flows for $p_T >$ 1.2 GeV/$c$. However, the $p_T$ range covered is not sufficient for a conclusive 
remark on the production mechanism of the $\rho^0$ measured in the hadronic decay channel. 
This measurement will improve with the STAR Time-of -Flight that will allow increasing  the $p_T$ range covered.

 % do not change 

\begin{thebibliography}{00} % do not change 
   
\bibitem{13} S. D. Protopopescu et al., {\it Phys. Rev. D} {\bf 7}  (1973) 1279.
\bibitem{14} M. J. Matison et al., {\it Phys. Rev. D} {\bf 9}  (1974) 1872.
\bibitem{15} K. Hagiwara et al., {\it Phys. Rev. D} {\bf 66} (2002) 010001.  
\bibitem{non} C. Nonaka et al., {\it Phys. Rev. C} {\bf 69} (2004) 031902.
\bibitem{26} X. N. Wang and M. Gyulassy, {\it Phys. Rev. D} {\bf 44} (1991) 3501; {\it Compt. Phys. Commun.} {\bf 83} (1994) 307.
\bibitem{16} J. Adams et al., {\it Phys. Rev. Lett.} {\bf 92} (2004) 92301.
\bibitem{pdg} C. Amsler et al., {\it Physics Letters B} {\bf 667} (2008) 1.
\bibitem{prc} J. Adams et al., {\it Phys. Rev. C} {\bf 71} (2005) 64902.
\bibitem{dau} B. I. Abelev et al., {\it Phys. Rev. C} {\bf 78} (2008) 44906.
\bibitem{v2} S. Shi for the STAR Collaboration, arXiv:0806.0763v2.
\bibitem{fit} X. Dong et al., {\it Phys.Lett. B} {\bf 597} (2004) 328.

\end{thebibliography}
\end{document}